\documentclass[useAMS,usenatbib]{mn2e}

\usepackage{hyphenat}
\usepackage{graphicx}
\usepackage{fixltx2e}
\usepackage{color}
\usepackage{amsmath}
\usepackage{amsfonts}
\usepackage{aas_macros}
\usepackage{dcolumn}
\usepackage{url}
\usepackage{natbib}

\newcommand{\revision}[1]{#1}
\newcommand{\ion}[2]{#1\textsc{#2}}
\newcommand{\set}[1]{\{#1\}}
\newcommand{\vecvar}[1]{\textbf{#1}}
\newcommand{\nn}{{N\!N}}

\newcolumntype{d}{D{.}{.}{-1} } 

\title
[Correlations of galaxy lines and continua]
{Quantifying correlations between galaxy emission lines and stellar continua}
\author[R. Beck, L. Dobos, C.W.~Yip, A.S.~Szalay and I. Csabai]{R\'obert Beck$^{1,2}$\thanks{E-mail:
beckrob23@caesar.elte.hu, dobos@complex.elte.hu, csabai@complex.elte.hu}, L\'aszl\'o Dobos$^{1}$, Ching-Wa Yip$^{2,3}$,
\newauthor
Alexander S. Szalay$^{2}$ and Istv\'an Csabai$^{1}$ \\
$^{1}$Department of Physics of Complex Systems, E\"{o}tv\"{o}s Lor\'{a}nd University, 1117 Budapest, Hungary \\
$^{2}$Department of Physics and Astronomy, The Johns Hopkins University, Baltimore, MD 21218, USA \\
$^{3}$Wolfram Research, Somerville, MA 02144, USA}
\begin{document}

\date{Accepted 2015 December 22. Received 2015 November 21; in original form 2015 June 4}

\pagerange{\pageref{firstpage}--\pageref{lastpage}} \pubyear{2015}

\maketitle

\label{firstpage}

\begin{abstract}

We analyse the correlations between continuum properties and emission line equivalent widths of star-forming and active galaxies from the Sloan Digital Sky Survey. Since upcoming large sky surveys will make broad-band observations only, including strong emission lines into theoretical modelling of spectra will be essential to estimate physical properties of photometric galaxies.  We show that emission line equivalent widths can be fairly well reconstructed from the stellar continuum using local multiple linear regression in the continuum principal component analysis (PCA) space. Line reconstruction is good for star-forming galaxies and reasonable for galaxies with active nuclei. We propose a practical method to combine stellar population synthesis models with empirical modelling of emission lines. The technique will help generate more accurate model spectra and mock catalogues of galaxies to fit observations of the new surveys. More accurate modelling of emission lines is also expected to improve template-based photometric redshift estimation methods. We also show that, by combining PCA coefficients from the pure continuum and the emission lines, automatic distinction between hosts of weak active galactic nuclei (AGNs) and quiescent star-forming galaxies can be made. The classification method is based on a training set consisting of high-confidence starburst galaxies and AGNs, and allows for the similar separation of active and star-forming galaxies as the empirical curve found by Kauffmann et al. We demonstrate the use of three important machine learning algorithms in the paper: $k$-nearest neighbour finding, $k$-means clustering and support vector machines

\end{abstract}


\section{Introduction}

Stellar population synthesis models are very successful in explaining the spectral energy distribution of galaxies in the optical \citep{Pegase1, BruzualCharlot2003, Maraston2011, Vazdekis2012} but they do not account for the characteristic emission lines originating from the excited interstellar gas. Starburst galaxies and galaxies with an active nucleus can produce emission lines so strong that can reach $~60$ per cent of the continuum flux in certain bands, or as much as 1 mag \citep{Atek2011}. As a result, pure population synthesis models are not enough to account for observations made with broad-band photometric filters. Since future large sky surveys will make photometric observations only, accurate modelling of the emission lines will be essential to estimate physical properties (including photometric redshifts) of galaxies precisely.

The purpose of this paper is to empirically quantify correlations between properties of the stellar continuum of galaxy spectra, and the strengths of emission lines. We also propose a recipe for generating realistic emission lines in the optical regime for stellar continua taken from population synthesis models. Moreover, we present a novel classification method to differentiate between starburst and active galaxies.

Results presented in the paper are obtained with the help of three important, widely used machine learning techniques that have just started to gain popularity in astronomical data analysis. Local linear regression using nearest neighbours \citep{Csabai2007, Kerekes2013} has been used for physical parameter estimation based on broad-band photometry. $k$-means clustering, an automatic, unsupervised classification algorithm has been applied successfully, for instance, to classify gamma-ray bursts \citep{Chattopadhyay2007, Veres2010}. Support vector machines (SVM), a supervised classification algorithm has been used for star--galaxy separation \citep{Kovacs2015} and transient detection \citep{Wright2015}. We will briefly introduce these methods later in the paper. For a detailed introduction to the field, refer to \citet{Ivezic2013}.

\revision{As with all training set-based empirical methods, the validity of our results is limited to the training set's coverage of the parameter space (in our case the redshift, metallicity, luminosity, continuum and line properties). Extrapolation capabilities of empirical techniques to parameter ranges outside the coverage is usually poor compared to theoretical models. While this certainly constrains the applicability of our results to strong emission line galaxies of the Sloan Digital Sky Survey (SDSS), the method itself can be easily extended to galaxies outside the investigated sample by augmenting the training set.}

The structure of the paper is as follows. In Sec.~\ref{sec:data}, we explain the sample selection and data reduction methods. Sec.~\ref{sec:emlines} describes the line reconstruction methods we investigated. \revision{An empirical method for star-forming--active galactic nucleus (AGN) separation is given in Sec.~\ref{sec:svm}. We present a stochastic procedure to generate realistic emission lines for continuum models in Sec.~\ref{sec:recipe}.} We summarize our findings and outline future work in Sec.~\ref{sec:summary}.

Wavelengths are generally quoted in vacuum. More information on the data used for this study, additional figures and program source code are available on the web site of the paper\footnote{\url{http://www.vo.elte.hu/papers/2015/emissionlines}}. Colour versions of the figures are available in the online version of the paper.

\subsection{Earlier work}

Thanks to the large amount of flux-calibrated optical galaxy spectra accumulated by the SDSS, precision of galaxy spectrum modelling has been improved significantly during the last decade. Many software tools and libraries exist to generate realistic stellar continua from a prescribed star formation history and various libraries of single stellar population spectra with a wide range of metallicities and initial mass function choices \citep{Pegase1, BruzualCharlot2003, Maraston2011, parsec, Vazdekis2012}. Models have also been extended with descriptions of interstellar extinction, the UV--IR balance \citep{grasil, CharlotFall2000, daCunha2010} and the chemical evolution of the gas from which stars can form \citep{Dave2011b}.

Emission lines of galaxy spectra carry a large amount of information about the abundance and ionization states of elements in the interstellar gas. Based on ionization ratios of the various elements, the source of primary radiation responsible for the excitation of the interstellar medium (ISM) can be characterized \citep{BPT1981, Kewley2001, Kauffmann2003a} The two primary radiation sources are young, hot, massive stars and active galactic nuclei (AGNs). Their different spectra (thermal and power law, respectively) cause different ionization states and ratios of the most common elements which, in turn, produce well measurable, strong, often broad emission lines: the Balmer series of hydrogen, [\ion{O}{ii}], [\ion{O}{iii}], [\ion{N}{ii}], [\ion{S}{ii}], etc. Population synthesis models do not account for the emission of the ISM.

\revision{Photoionization models \citep{Stasinska1984, cloudy} yield accurate line ratios for any primary radiation spectrum and gas composition. To couple stellar population synthesis with models of photoionization, shock-heating of the interstellar gas, emission of the dust etc., the star formation history, several interactions between the stellar populations, the active nucleus, the dust and gas content need to be accounted for. For instance, AGNs are very likely to be responsible for quenching rapid star formation following starburst periods in the galaxy but they also emit ionizing radiation that excites gas, evaporates dust and produces shock waves that heat the ISM. Also, starburst periods are followed by high supernova activity that enriches the ISM with metals, leading to significant chemical evolution which must be reflected in the models of emission lines. Additionally, a recent advancement in stellar population synthesis is the inclusion of stellar rotation and binary evolution effects which have been shown to noticeably influence the strengths of some emission lines \citep{Eldridge2012, Stanway2014, Leitherer2014, Topping2015}. Taking everything into account is not possible without detailed hydrodynamic simulation of the galaxies \citep{Jonsson2010, Kewley2013} or without making significant simplifications to the models. Various software, notably P\'{E}GASE and BPASS \citep{Pegase1, Eldridge2012}, can be used to generate emission lines on top of stellar continua computed from stellar population synthesis. The photoionization part of these softwares, however, introduces a large set of free parameters that describe the distribution and composition of the ISM. A frequently used way of reducing the number of free parameters is to make theoretical or empirical assumptions. Typical theoretical simplifications include the assumption of spherical symmetry or the use of a common ionization spectrum for all gas clouds \citep{Stasinska1984, Pegase1, cloudy}. If no strict physical considerations can be made, to generate realistic emission lines on top of \revision{modelled} continua using any photoionization code, one has to estimate the \textit{a priori} distribution of model parameters by comparing large ensembles of models with observations. For instance, the code Le Phare \citep{LePhare} uses the relations of \citet{Kennicutt1998} to parametrize emission lines.}

Another route to take to generate realistic emission lines is to work on an entirely empirical basis. \citet{Yip2004} demonstrated that stellar continua of SDSS galaxies form a 1D sequence and thus, can be characterized by a single numerical value, the \texttt{eclass}. The value of \texttt{eclass} for each galaxy spectrum is obtained by expressing the continuum on a basis derived from principal component analysis (PCA). \citet{Gyory2011} showed that strong correlations between the \texttt{eclass} (i.e. the stellar continua) of starburst and AGN galaxies exist. They applied PCA to expand emission line equivalent widths (EWs) of SDSS galaxies on a 3D basis and correlate the principal components with the \texttt{eclass} of the continua. We take their approach a step further: based on the correlations, we give a recipe to automatically generate emission lines with realistic distribution and refine star-forming--AGN separation using the principal components and SVM.

\section{Data reduction}
\label{sec:data}

We started with the entire spectroscopic galaxy sample of the Sloan Digital Sky Survey Data Release 7 \citep{SDSS DR7; Abazajian2009} which we later filtered by signal-to-noise ratio and line strength. As one of our goals was to accurately fit broad AGN lines, we measured line parameters ourselves.

\subsection{Continuum fitting and line measurements}
\label{section_fits}

PCA is widely used to derive a representative basis from optical spectra of galaxies. When performing PCA on emission line galaxies, the eigenspectra are primarily sensitive to the variations in emission line strengths and only secondly to continuum features \citep{Connolly1995, Yip2004}. \revision{Obviously, the} slope of the continuum is correlated with emission lines but the variance of the lines is bigger. To run PCA on the pure continua, one has to mask the regions of emission lines, or eliminate the lines completely by subtracting line models from the measured spectra. Line fits have to be precise enough so that the line-subtracted continua contain minimal residuals. We reprocessed the entire set of SDSS DR7 galaxy spectra according to these requirements with our own implementation of the algorithm detailed in this section.

One frequent method of fitting continua in the optical band is to express the spectrum as a non-negative linear combination of template spectra \citep{tremonti2004} while also accounting for the intrinsic attenuation and velocity dispersion. Although more advanced, Bayesian and PCA-based methods exist \citep{Kauffmann2003b, chen2012} to derive physical properties from the continuum, as we were mainly interested in the emission lines, we retained the former technique for continuum subtraction. First, we corrected for galactic extinction, masked emission lines and fitted the continuum using the templates from \citet{BruzualCharlot2003} by also fitting the velocity dispersion and intrinsic extinction in parallel. Intrinsic extinction was modelled following \citet{CharlotFall2000}. Metallicity was taken into account by fitting \revision{four} sets of templates \revision{of differing metallicities} and choosing the one with minimal reduced $\chi^2$. \revision{Thus, the fitted metallicity can take one of four values: $Z=0.004$, $0.008$, $0.02$ or $0.05$.} \revision{We did not take the nebular continuum emission into account, which, in the case of young starburst galaxies, can contribute a non-negligible flux to the near-infrared part of the spectrum \citep{Leitherer1995}. Since the entire continuum was fitted with stellar templates only, we expect a slight overestimation of absorption lines, and therefore the overestimation of emission lines for starburst galaxies. On the other hand, within the wavelength coverage of SDSS spectroscopy, nebular continuum emission is significant only in the case of stellar populations younger than $10$~Myr or at very low metallicities of $Z \sim 0.0001$ \citep{Molla2009}, and only about 0.5~per~cent of our sample potentially fall into this parameter range.}

\begin{figure*}
\begin{minipage}[t]{\textwidth}
	\includegraphics{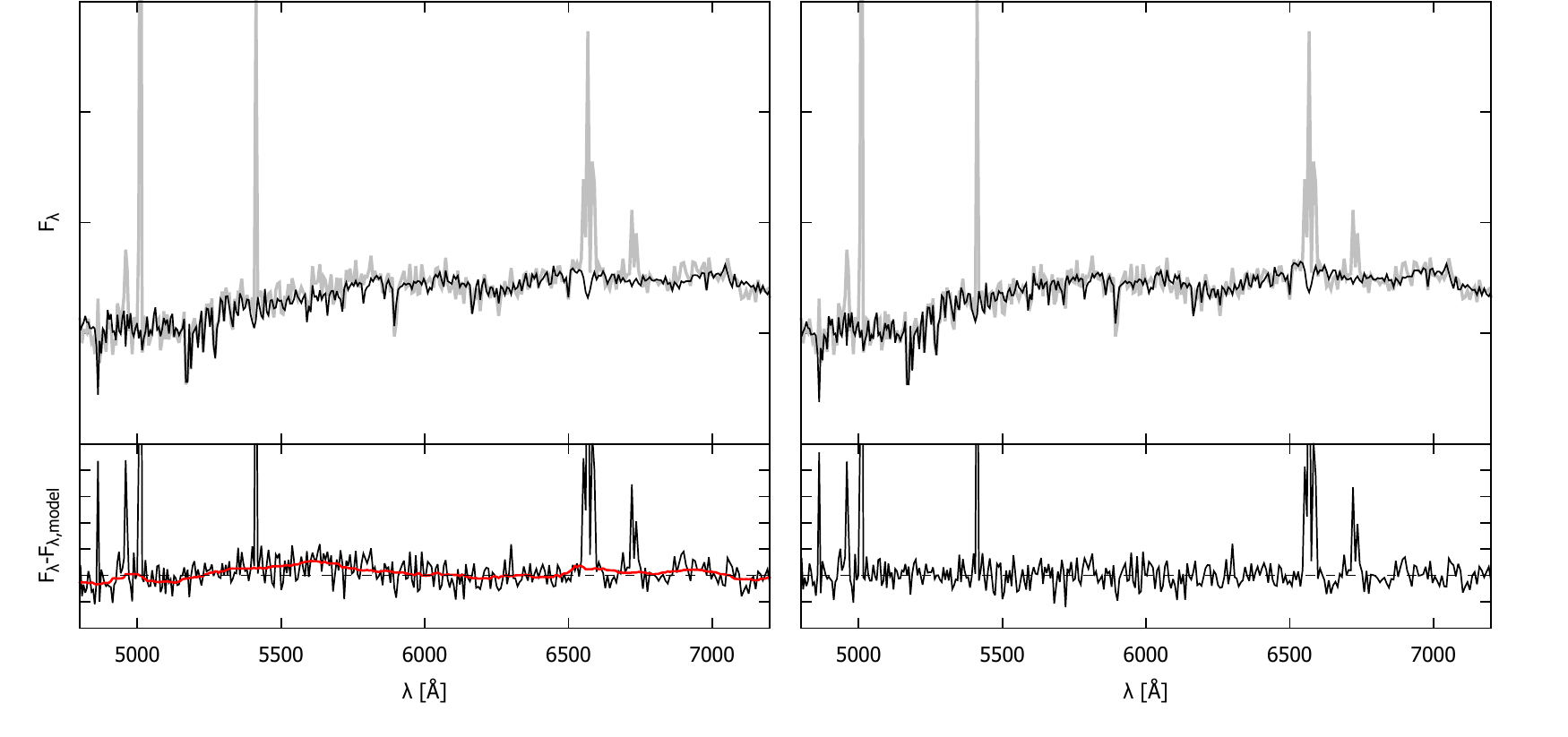}
	\caption{Illustration of fitting the stellar continuum. In the top left panel the best non-negative least square fit from 10 Bruzual--Charlot templates is plotted, the residual is visible in the bottom left panel. The effect of the low-pass filter on the residual is drawn with a red curve in the bottom left panel; we subtract this curve from the noisy residual prior to fitting emission lines. The top right panel illustrates the best-fitting continuum model, corrected for discrepancies by adding back the low-pass-filtered residual to the stellar population synthesis spectrum. The top right panel shows the high-pass-filtered residual used for fitting the lines.}
	\label{fig:contfit}
\end{minipage}
\end{figure*}

Due to discrepancies between continuum models and SDSS spectra \citep{Maraston2009}, the continuum-subtracted spectrum consists of three components: the emission lines, the noise and a slowly changing background that originates from the imperfect models. Since the emission lines and noise are high-frequency components, one can easily eliminate the background by a high-pass filter. For this purpose, we used a 50~\AA~wide rolling median filter. This was wide enough to leave broad AGN lines almost intact, yet remove any residuals of the incorrect background subtraction. Fig.~\ref{fig:contfit} illustrates this procedure.

\begin{figure*}
\begin{minipage}[t]{2.0\columnwidth}
	\hspace{-0.5cm}
		\includegraphics{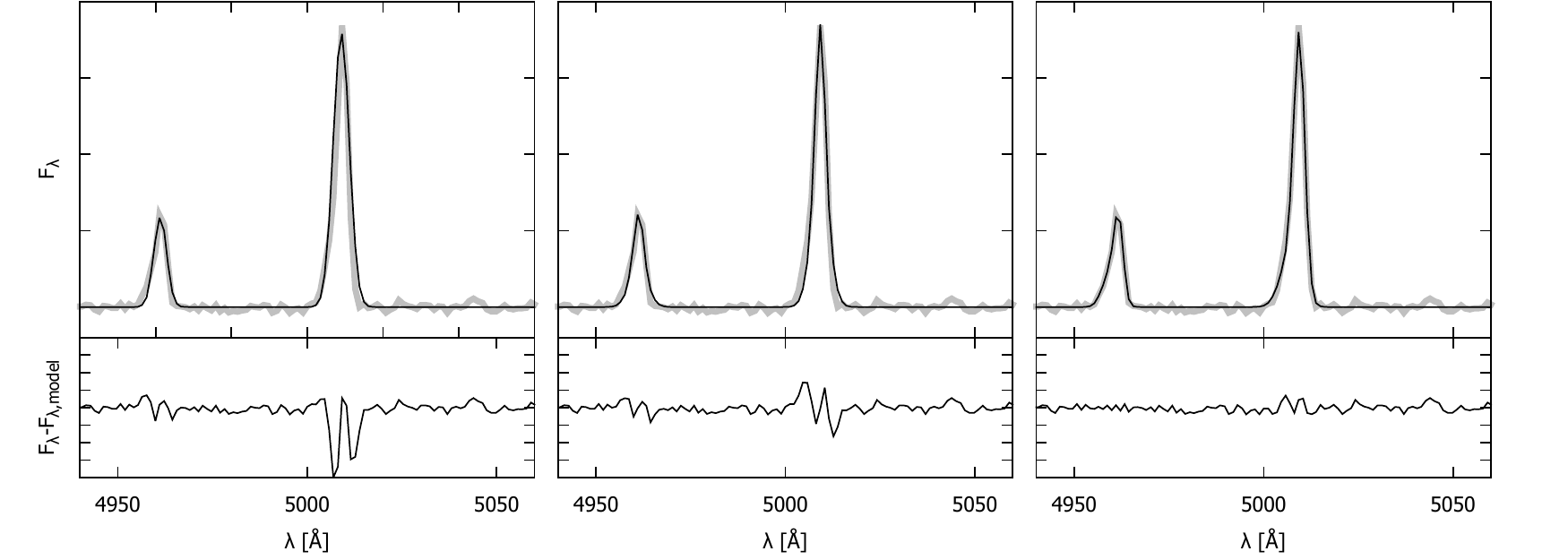}
		\caption{Illustration of noise-limited fitting of asymmetric emission lines with increasingly complex models. The top panels show the original continuum-subtracted spectrum in grey and the best-fitting models in black. The bottom panels show the residuals. The left-hand panel corresponds to a single Gaussian fit, the middle panel to two Gaussians centred on the same mean wavelength but with different variance while the right-hand panel shows the results from fitting two Gaussians with slightly different centre wavelengths. In this case, the most complex model is accepted as the line residuals are higher than the average noise for both simpler models, whereas the line residual is comparable to the average noise in the third case.}
		\label{fig:linefitting}
\end{minipage}
\end{figure*}

Once the low-frequency background \revision{has been} removed, lines are fitted using a technique we call \textit{noise-limited fitting}. To precisely fit all strong emission lines, including those of active galaxies, we use three increasingly complex line models.
\begin{itemize}
\item A single Gaussian:
	\begin{equation*}
		F(\lambda) = A \cdot \mathrm{e}^{- \frac{(\lambda - \lambda_0)^2}{\sigma^2}}
	\end{equation*}
\item Two Gaussians centred on the same wavelength but with different 
variance
	\begin{equation*}
		F(\lambda) = A \cdot \mathrm{e}^{- \frac{(\lambda - \lambda_0)^2}{\sigma_a^2}} + B \cdot \mathrm{e}^{- \frac{(\lambda - \lambda_0)^2}{\sigma_b^2}}
	\end{equation*}
\item Two Gaussians allowing for a small offset $\Delta \lambda < 5$~\AA\ between the centres, different variance
	\begin{equation*}
		F(\lambda) = A \cdot \mathrm{e}^{- \frac{(\lambda - \lambda_a)^2}{\sigma_a^2}} + B \cdot \mathrm{e}^{- \frac{(\lambda - \lambda_b)^2}{\sigma_b^2}}
	\end{equation*}
\end{itemize}
While the first model is enough to fit emission lines with typical velocity dispersion, the second model is necessary for lines with broad wings and the third model for asymmetric lines. Our objective is to find the simplest, yet well-fitting model. Overlapping emission lines are -- obviously -- fitted together, but we do not enforce any correlation on the EWs of lines from the same ion. \revision{Also, the velocity dispersions of the lines, even of those from the same ion, are fitted independently.} First, we fit the lines with the simplest model, subtract it from the measurement and compare the residual within the region of the emission line with the noise in wavelength ranges without lines. If rms of the residual inside the region of the line is at least two times than elsewhere, we reject the model and attempt to fit the line with a more complex one. Fig~\ref{fig:linefitting} illustrates how this technique works on asymmetric broad AGN lines.

\begin{figure}
	\includegraphics{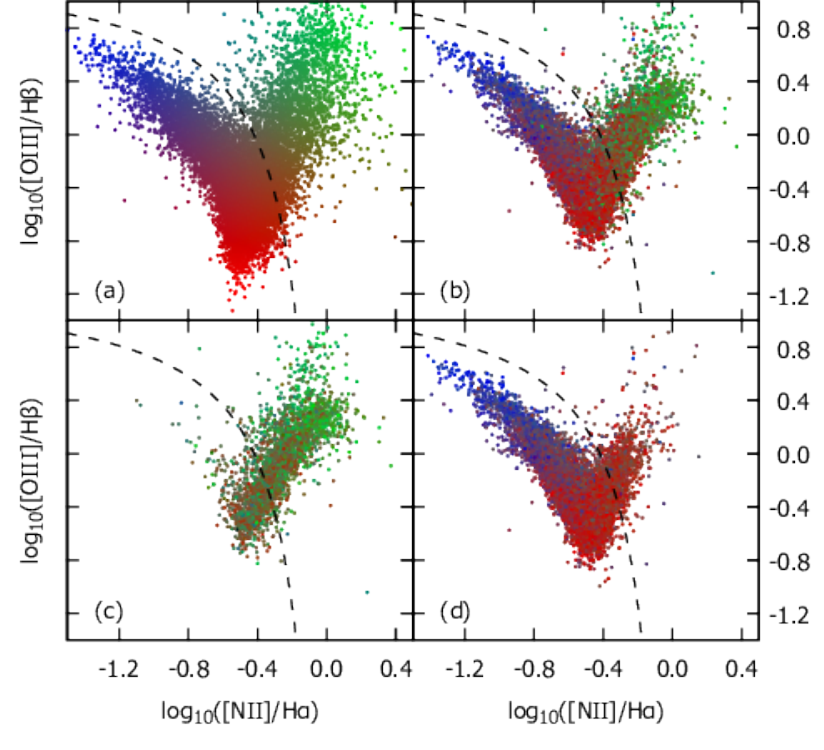}
	\caption{Original and reconstructed BPT diagrams of strong emission line galaxies sampled from SDSS~DR7. In each panel, the dashed curved shows the empirical segregation line between star-forming galaxies and AGN as defined by \citet{Kauffmann2003a}. Panel (a) is plotted from directly measured line EWs. Galaxies are colour coded based on their loci in the BPT plane: blue galaxies are star-forming, green ones are AGNs and red ones are the intermediate weak AGNs in the bottom corner of the distribution. This colour coding based on directly measured emission lines is used in all BPTs throughout the paper. Panel (b) displays the BPT of line log~EWs reconstructed from continuum principal components using the local linear regression method with the 30 nearest neighbours in PCA space. Panels (c) and (d) show galaxies only that were originally classified as (c) AGNs, (d) star-forming using directly measured line EWs. While lines of strong AGNs and extreme starburst galaxies can be reconstructed well, there is significant `cross-talk' in the quiescent region.}
	\label{fig:bpt_classes}
\end{figure}

Tab.~\ref{table_lines} summarizes the fitted and subtracted emission lines. Line model fit parameters are available online. In panel (a) of Fig.~\ref{fig:bpt_classes}, we plot the Baldwin--Phillips--Terlevich (BPT) diagram of the sample using a colour coding we are going to use throughout the paper.

\begin{table}
\label{table_lines}
\begin{center}
\begin{tabular}{l l | l l | l l}
	\hline
	Line & $\lambda_\textnormal{vac}$ (\AA) & 
	Line & $\lambda_\textnormal{vac}$ (\AA) &
	Line & $\lambda_\textnormal{vac}$ (\AA) \\ \hline \hline
	\ion{O}{ii} & 3727.09  &  	  H$\gamma$ & 4341.68 &	  			\ion{O}{i} & 6365.54 \\ 
	\ion{O}{ii} & 3729.88  &	  \ion{O}{iii} & 4364.44 &          \ion{N}{i} & 6529.03 \\ 
	H$\theta$ & 3798.98  &		  H$\beta$ & 4862.68 &	            \ion{N}{ii} & 6549.86 \\
	H$\eta$ & 3836.47 &			  \ion{O}{iii} & 4932.60 &          H$\alpha$ & 6564.61 \\  
	H$\zeta$ & 3890.16 & 		  \ion{O}{iii} & 4960.30 &          \ion{N}{ii} & 6585.27 \\
	H$\epsilon$ & 3971.20 &		  \ion{O}{iii} & 5008.24 &          \ion{S}{ii} & 6718.29 \\
	\ion{S}{ii} & 4072.30 &		  \ion{He}{i} & 5877.65 &           \ion{S}{ii} & 6732.67 \\
	H$\delta$ & 4102.89 &		  \ion{O}{i} & 6302.05 & &   	\\
	\hline
\end{tabular}
\caption{List of the fitted nebular emission lines.}
\end{center}

\end{table}

\subsection{Comparison with other work}

It is interesting to compare our line fits to those of \citet{Brinchmann2004}. In the cited work, the authors used a simpler technique of fitting nebular emission lines of SDSS galaxies with the primary focus on the signal-to-noise ratio of line measurements and not on the minimization of the residuals after line subtraction. As a result, their line models cannot directly be used to get a pure continuum due to the high residuals of the fitting.

\begin{figure*}
	\vspace{8pt}
	\hspace{8pt}
	\includegraphics{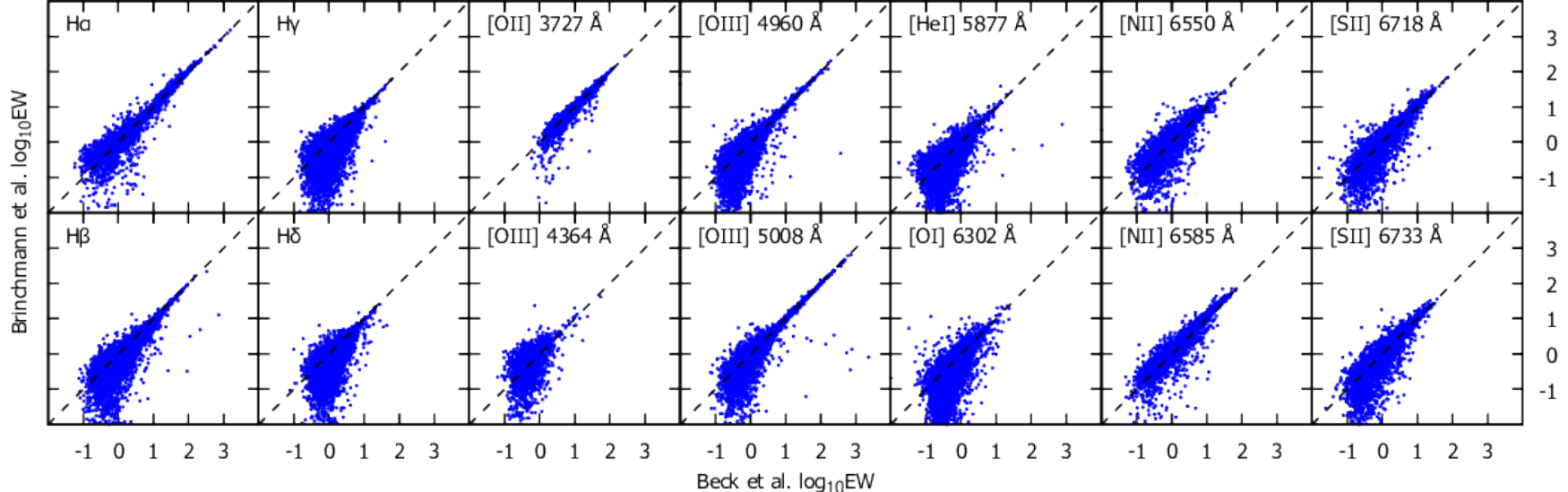}
	\caption{Comparison of emission line equivalent width of \citet{Brinchmann2004} ($y$-axis) with ours ($x$-axis). EWs are expressed in angstroms; scales on both axes are the same. Density plots are normalized for EW bins in such a way that stronger lines are also visible. Our estimate on \revision{weak lines is systematically higher, which appears more pronounced due to the log scale, but is not that significant compared to errors.}}
	\label{fig:mpacompare}
\end{figure*}

In Fig.~\ref{fig:mpacompare}, we compare the EWs of the most prominent emission lines as derived with our technique and with the method of \citet{Brinchmann2004}. \revision{In the case of strong emission lines, our measurements of line strengths are very similar to the results of \citet{Brinchmann2004}, but we estimate weak emission lines significantly higher. This is very likely due to the high-pass filtering applied to the continuum-subtracted spectrum, cf. Sec.~\ref{section_fits}. Yet, between $97.9\%$ and $99.1\%$ of our line, EWs are within $3\sigma$ of \citet{Brinchmann2004}, with the exception of H$\alpha$ and H$\beta$ where only $93.9\%$ and $87.7\%$, respectively, of the measurements are within $3\sigma$. Also, \citet{Brinchmann2004} measured weak lines by fitting them together with stronger lines of the same ion, imposing a constraint on line ratios, whereas we fitted these lines independently. Weaker lines can easily become undetectable in noisy regions, hence our fitting method introduces some selection bias.}

\subsection{Galaxy sample selection}

We selected a smaller sample of $N = 13788$ galaxies from the entire set of continuum and line-fitted spectra that met the following criteria:

\begin{itemize}
	\item observed at a signal-to-noise ratio $S/N > 5$,
	\item all 11 emission lines listed in Tab.~\ref{tab:ewfit} are measured and non-zero. These lines are the same as in \citet{Gyory2011}.
\end{itemize}

\revision{The sample size was further limited to an easily manageable number by choosing a section of the sky (right ascension between $220^{\circ}$ and $230^{\circ}$).}

\begin{figure}
	\includegraphics{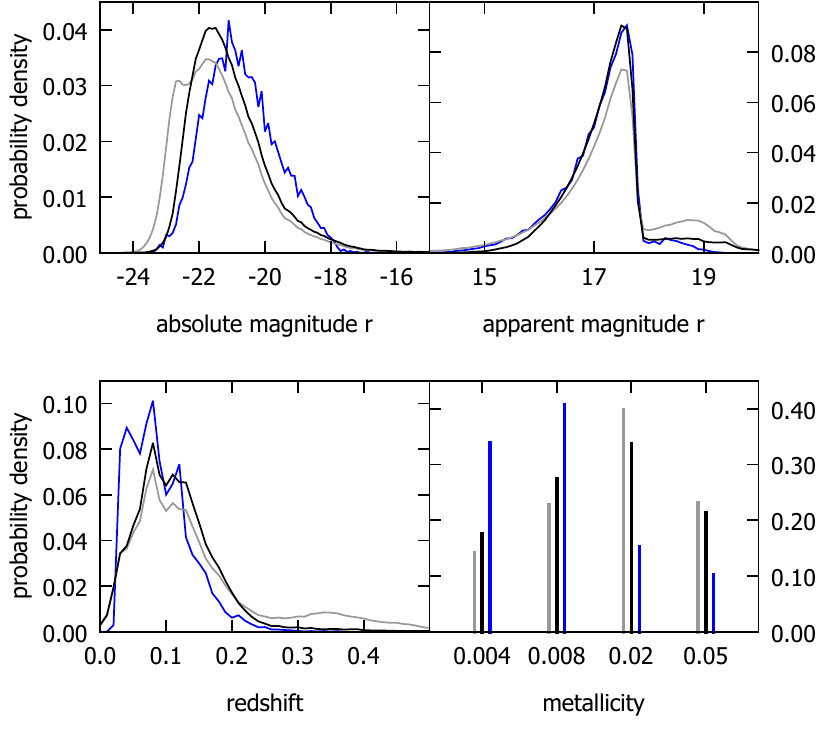}
	\caption{\revision{Normalized histograms of galaxy properties. Our sample is plotted in blue, the grey lines correspond to the entire DR7 spectroscopic galaxy sample, while the black lines show the DR7 sample excluding the deeper LRG sub-sample. The latter provides a better comparison to our data set, since we selected predominantly from the main galaxy sample.}}
	\label{fig:samplehistograms}
\end{figure}

\revision{The requirement that all $11$ emission lines should be measurable results in a sample containing galaxies with ongoing star formation or possessing an active nucleus only. Fig.~\ref{fig:samplehistograms} shows the selection effects on the distribution of the apparent and absolute $r$-band magnitudes, the redshift and the metallicity. While the cut in signal-to-noise ratio did produce a cutoff around $r=19$ apparent magnitude and a relative increase of objects towards smaller redshifts, the absolute magnitude distribution shows that our selection method prefers fainter, smaller and younger galaxies. Galaxies with lower metallicities are also selected with higher probability, presumably due to correlations between ongoing strong star formation, metallicity and age. Nevertheless, galaxies with solar and above solar metallicities are still present in the sample.}

\revision{The SDSS~DR7 main galaxy sample, which makes up the majority of our training set, was not selected for morphological type or colour, and thus includes a wide variety of galaxies \citep{Strauss2002}. At larger redshifts, however, different environments, e.g. harder radiation fields and higher ionization parameters \citep{Steidel2014}, can lead to significantly different emission line characteristics. Certainly, the validity of our results is constrained by parameter ranges covered by the sample. By using a data set that goes beyond the types of galaxies observed by the SDSS, one can easily apply our method to a broader range of galaxies.}

\subsection{Continuum principal components}
\label{sec:contpca}

Principal components of the stellar continuum were derived from the fitted model spectra instead of the measurements directly. Although the precise line modelling would make it possible to subtract emission lines from the original spectra or run PCA directly on the measurements by masking out emission lines, due to the limited size of the sample which would make eigenspectra noisy, we choose to use the models instead. Fitted continuum models were taken at rest frame, convolved with the best-fitting velocity dispersion kernel and normalized to have equal flux in the following featureless rest-frame wavelength ranges: $4250$--$4300$~\AA, $4600$--$4800$~\AA, $5400$--$5500$~\AA, $5600$--$5800$~\AA. PCA was done in the $3722$--$6761$~\AA{ }range with $0.6$~\AA{ }binning. The average continuum was subtracted from the individual spectra prior to calculating the covariance matrix.

Eigenspectra were determined using the Lanczos singular value decomposition (SVD) algorithm from PROPACK \citep{propack}. The algorithm calculates only a given number of singular vectors with the largest corresponding singular values. This was very useful in our case as the spectra consisted of $5065$ data points, whereas we were interested in the first five principal components only.

\begin{figure}
	\includegraphics{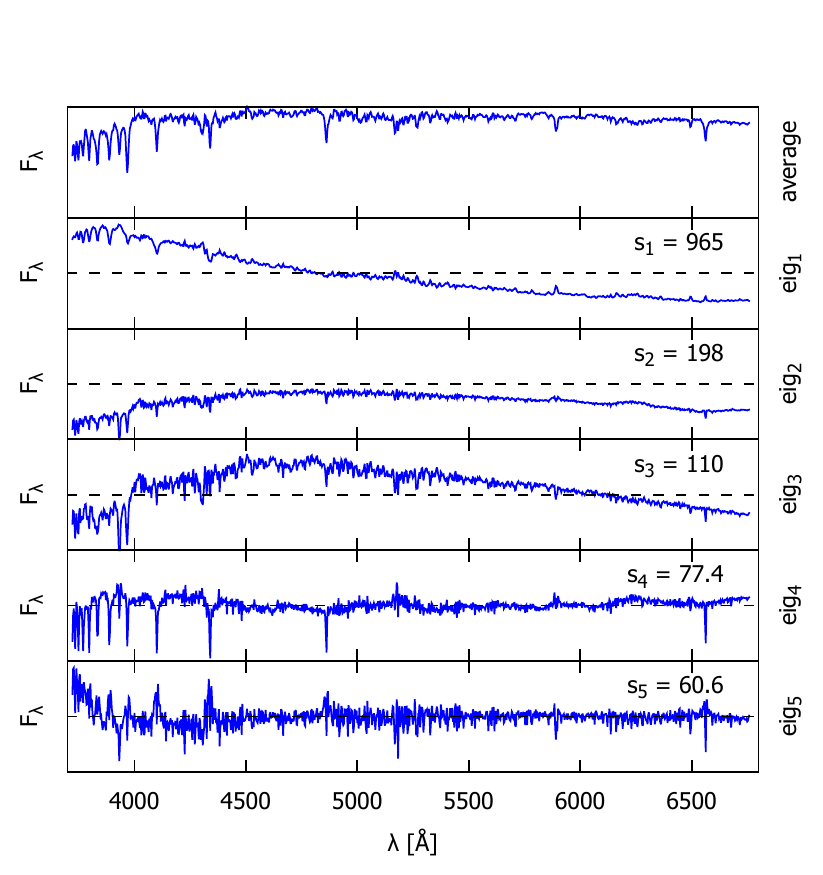}
	\caption{The average and the first five eigenvectors of the principal component analysis of galaxy continua, ordered by the corresponding singular values (as displayed in each panel). See the text for the physical interpretation of the eigenspectra.}
	\label{fig:pcavectors}
\end{figure}

The average spectrum and the resulting eigenspectra are plotted in Fig.~\ref{fig:pcavectors}. As the average was subtracted, the first eigenspectrum corresponds to the colour of the galaxy. The following two basis vectors are rather similar at first sight but the third one shows more prominent absorption lines. They are together very likely to determine the age and metallicity of the galaxy as the $4000$~\AA-break is very strong in both of them. The fourth vector probably corresponds to the width of absorption lines thus correlates with velocity dispersion. \revision{The magnitude of the fourth and fifth eigenvalues is similar, and they already mark the start of the plateau in the distribution of eigenvalues, therefore taking more eigenspectra into account does not significantly increase the variance explained by them.}

\subsection{Emission line principal components}

\begin{figure}
	\includegraphics{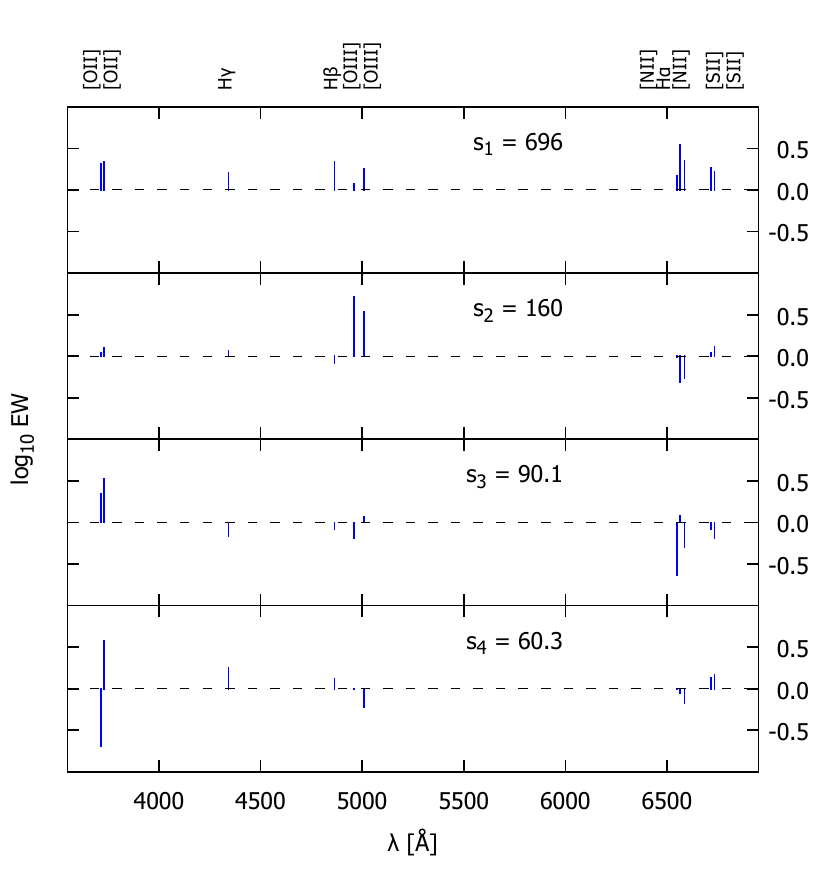}
	\caption{The first four singular vectors of the correlation matrix of the logarithm of emission line equivalent widths, ordered decreasingly by the corresponding singular values (as displayed in each panel). The fourth vector is very likely to be just noise as [\ion{O}{ii}] lines should not have different signs.}
	\label{fig:linepcavectors}		
\end{figure}

In contrast to what was done by \citet{Gyory2011}, we calculate principal components of the \textit{logarithm} of emission line EWs. Fig.~\ref{fig:linepcavectors} shows the resulting singular vectors. Taking the logarithm is more useful when one is interested in line ratios instead of absolute line strengths and uses linear methods for the analysis. We have to mention, however, that using the logarithm of the EWs also means that the results presented in the rest of the paper will be valid in the \textit{logarithmic sense} only.

\section{Reconstructing emission lines}
\label{sec:emlines}

Our goal was to empirically estimate emission line EWs from continuum principal components. If there exists any correlation between the continuum and emission lines of galaxy spectra, it is clearly non-linear. Global linear methods to analyse the correlations are not useful in this case, yet \textit{locally linear} methods still can be used.

\subsection{Local linear regression}
\label{sec:linreg}

Let us consider an ensemble of measurements where measured values are split into two sets $D = \set{\vecvar{d}_i}$ and $R = \set{r_i}$, $i$ indexing the individual measurements. For the sake of simplicity, $r_i$ are taken to be scalars whereas $\vecvar{d}_i$ are vectors, thus $D$ forms a metric space of dimension $N$. The Euclidean metric is often used to measure distances among data vectors of $D$ even though it might lack any physical interpretation. Our objective is to characterize known, or predict unknown $r_i$ from the always known $\vecvar{d}_i$ vectors. To estimate $r_i$ from $\vecvar{d}_i$, first we find the $k$-nearest neighbours of $\vecvar{d}_i$ in $D$. Let us denote the set of indices of these nearest neighbours with $\nn(\vecvar{d}_i, D, k)$, where $i \notin \nn $ by definition. Then we express $r_i$ in the following form
\begin{equation}
	r_i \approx c_i + \vecvar{a}_i \vecvar{d}_i.
	\label{eq:lincomb}
\end{equation}
Note, that both $\vecvar{a}_i$ and $\vecvar{d}_i$ are vectors and their dot product is taken in the formula above. The $c_i$ constants and the $\vecvar{a}_i$ coefficients need to be determined individually for every $(\vecvar{d}_i, r_i)$ measurement using standard linear regression by minimizing
\begin{equation}
	\chi_i^2 = \sum_{j \in \nn} \frac{\left(r_j - c_i - \vecvar{a}_i \vecvar{d}_j \right)^2}{w_j},
	\label{eq:chisquared}
\end{equation}
where $i$ is still the index of the measurement, $j$ runs on the nearest neighbours and $w_j$ is a weight. The expression of $\chi^2$ is similar if $r_i$ are vectors instead of scalars but the $\vecvar{a}$ coefficients become matrices. Errors in $r_i$ and the components of $\vecvar{d}_i$ can be incorporated into the value of $w_i$. Similarly, neighbours in $\nn$ can be ordered by distance from $d_i$ and the inverse of (the square of) the distance can be used as a weight in Eq.~\ref{eq:chisquared}.

Local linear regression has many advantages over global non-linear modelling. First of all, global models are usually either too simple to describe the data or prone to overfitting. Local linear models, on the other hand, are simple and can be used to characterize the local estimation errors. For instance, one can measure the goodness of the estimation of $r_i$ by the $\chi^2$ of the local linear fit. The challenge in local linear fitting is to find the $k$-nearest neighbours quickly in large data sets. Spatial indexing, most often a $kD$-tree index is used for this purpose \citep{Csabai2007}.

\subsection{Emission line reconstruction from the continuum}
\label{sec:pcafit}

We applied the local linear regression technique to estimate emission line EWs from the stellar continuum principal \revision{components}. To test whether continuum PCs carry more information regarding the emission lines than broad-band SDSS magnitudes, we will also perform the regression analysis directly on the photometric magnitudes in Sec.~\ref{sec:magfit}. Further tests are done with randomized samples (Sec.~\ref{sec:randomfit}) to get a picture of the performance of our method.

By using the notation of Sec.~\ref{sec:linreg}, $\vecvar{d}_i$ became the first five continuum principal components and $r_i$ became the log~EWs. Emission line log~EWs were fitted individually based on the log~EWs of the $k=30$ nearest neighbour galaxies in the continuum PCA space. The $\chi^2$ of the fitting was weighted by the inverse-square distance of the neighbours from the query point. \revision{The value of $k=30$ was chosen as a rule of thumb: we are fitting $5+1$ parameters and the number of data points must be large enough to adequately determine that many parameters but small enough to preserve locality. Modifying this parameter within reasonable limits (e.g. $25-40$) does not significantly impact the results.}

\begin{figure}
	\includegraphics{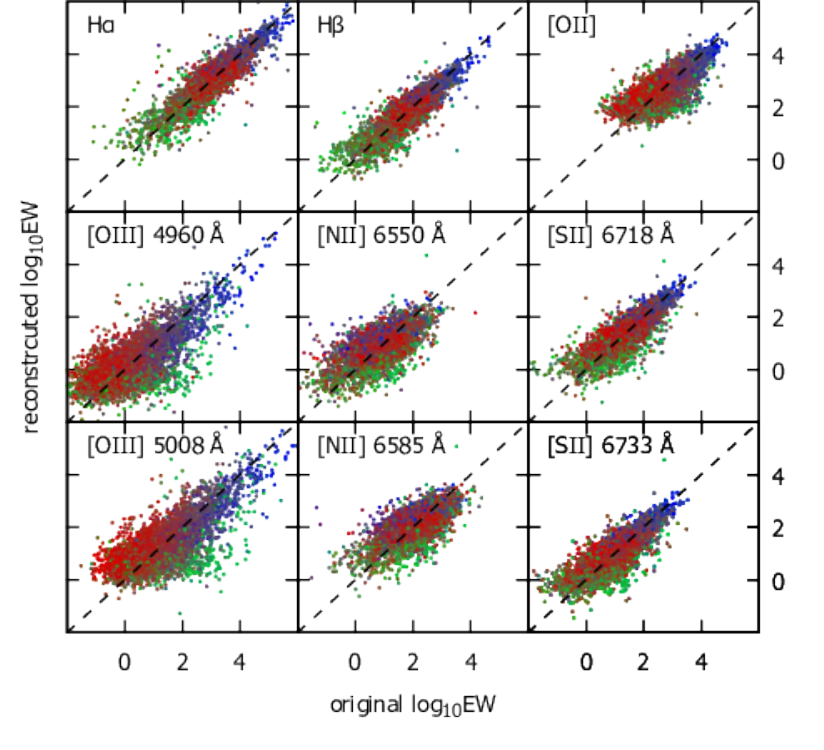}
	\caption{Reconstructed log~EWs from continuum principal components. Estimated log~EWs are plotted as functions of the directly measured log~EWs for the $11$ emission lines we used. Colour coding of data points is the same as in panel (a) of Fig.~\ref{fig:bpt_classes} and reflects the activity class of galaxies.}
	\label{fig:ewfit}		
\end{figure}

\begin{figure}
	\includegraphics{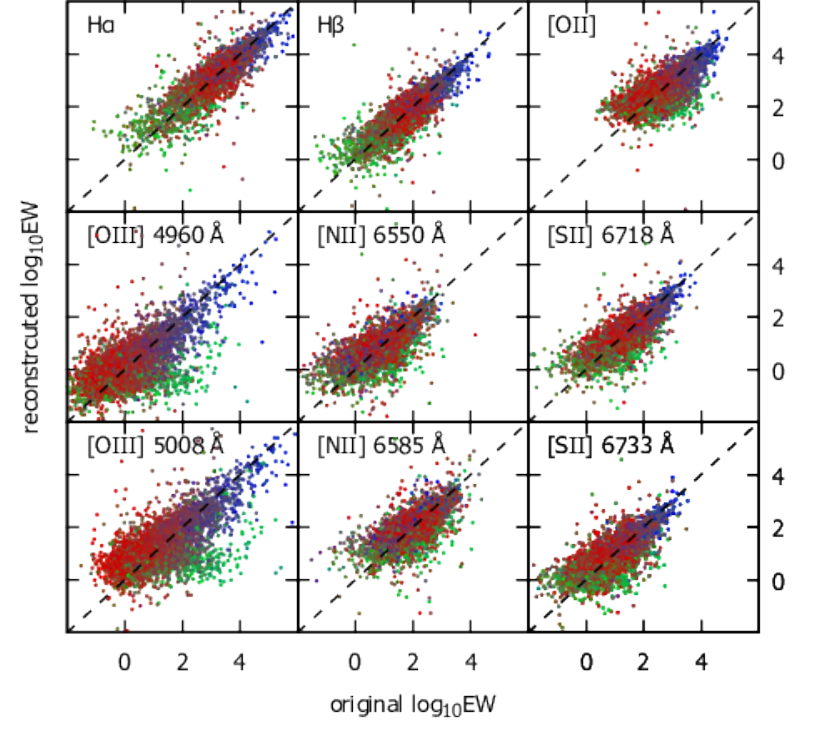}
	\caption{Reconstructed log~EWs from broad-band SDSS magnitudes. Estimated log~EWs are plotted as functions of the directly measured log~EWs for the $11$ emission lines we used. Colour coding of data points is the same as in panel (a) of Fig.~\ref{fig:bpt_classes} and reflects the activity class of galaxies.}
	\label{fig:magfit}
\end{figure}

\revision{Fig.~\ref{fig:ewfit} shows the reconstructed log~EWs of emission lines as functions of the directly measured EWs. EWs reconstructed from the continuum are in reasonably good agreement with directly measured log~EWs. The relative flux error $\sigma_r$ of the line reconstruction is Gaussian but a systematic shift $\delta$ is visible in the case of [\ion{O}{ii}], [\ion{O}{iii}] and [\ion{N}{ii}] ($\delta \approx 0.1, 0.15, 0.15$, respectively). The typical value of the relative error is $\sigma_r \approx 0.3$ for hydrogen and sulfur, $\sigma_r \approx 35\%$ for [\ion{O}{ii}] and [\ion{N}{ii}], and $\sigma_r \approx 45\%$ for [\ion{O}{iii}].}
\revision{Col.~3-4} of Tab.~\ref{tab:ewfit} list the outcome of the correlation analysis for the 11 investigated lines using the local linear regression technique. Pearson's product-moment correlation coefficient $\rho$ and the rms error $\sigma$ were calculated for each line. \revision{These numbers also show that} fits are most accurate for the hydrogen and sulfur lines ($\rho > 0.8$) whereas oxygen and nitrogen lines are significantly less correlated with direct line measurements. 

\begin{figure}
	\includegraphics{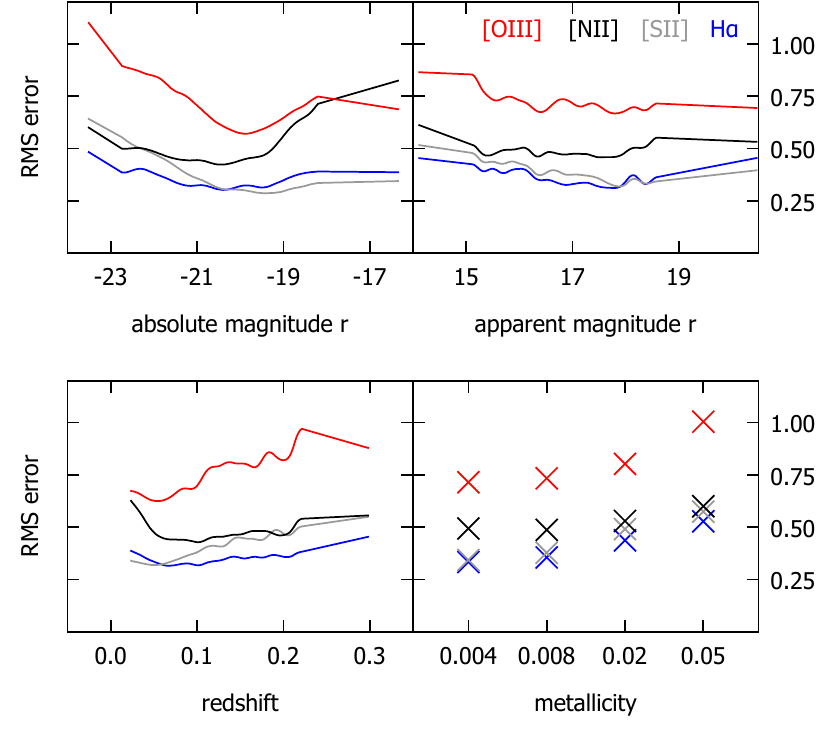}
	\caption{\revision{The rms error of emission line log~EW reconstruction as the function of various galaxy properties. The colours correspond to the following emission lines: red -- {[\ion{O}{iii}] 5008~\AA}, black -- {[\ion{N}{ii}] 6585~\AA}, grey -- {[\ion{S}{ii}] 6718~\AA}, and blue -- H$\alpha$. See the text for a discussion.}}
	\label{fig:sliceerrors}
\end{figure}

\revision{Fig.~\ref{fig:sliceerrors} shows the dependence of the error of the reconstruction on galaxy properties (cf. Fig.~\ref{fig:samplehistograms} for histograms of these) for select lines. The brighter and higher metallicity galaxies have a larger fraction of AGNs and are estimated with higher errors, especially in the case of oxygen and sulfur lines. Objects at higher redshifts generally exhibit decreasing accuracy. The error of line reconstruction visibly increases towards the limits of our training set. This is due to the fact that near the edges of the training set there are fewer galaxies and the nearest neighbours used to estimate the emission lines are generally less similar to each other and to the galaxy whose lines are being fitted.}

As we expected, emission lines can be much better reconstructed from the continuum of star-forming galaxies due to the strong connection between the young stellar population and the ISM: young massive stars are responsible for the excitation of interstellar gas clouds. Nevertheless, [\ion{O}{ii}] and [\ion{N}{ii}] lines show a significant scatter even in the star-forming case. Interestingly, sulfur lines can be reconstructed much better.

One intriguing result is that, while [\ion{O}{iii}] is an important indicator of nuclear activity, its reconstruction from continuum properties in case of AGNs seems rather problematic. It is understandable as AGN activity correlates much less with the properties of the stellar populations than in the star-forming case. Yet, some connections exist as it is visible from [\ion{N}{ii}] and the hydrogen lines.

\begin{table*}
	\begin{tabular}{l c | d d | d d | d d | d d }
		\hline
		& 
		& \multicolumn{2}{c|}{Principal components} 
		& \multicolumn{2}{c|}{Magnitudes} 
		& \multicolumn{2}{c|}{Principal components} 
		& \multicolumn{2}{c}{Randomized} \\
		
		& 
		& \multicolumn{2}{c|}{30 $\nn$ galaxies} 
		& \multicolumn{2}{c|}{30 $\nn$ galaxies}	 
		& \multicolumn{2}{c|}{30 random galaxies}
		& \multicolumn{2}{c}{30 $\nn$ galaxies} \\
		
		line 
		& $\lambda$ [\AA] 
		& \multicolumn{1}{l}{$\rho$}  
		& \multicolumn{1}{l|}{$\sigma$}	
		& \multicolumn{1}{l}{$\rho$}
		& \multicolumn{1}{l|}{$\sigma$}
		& \multicolumn{1}{l}{$\rho$} 
		& \multicolumn{1}{l|}{$\sigma$} 
		& \multicolumn{1}{l}{$\rho \times 10^2$}
		& \multicolumn{1}{l}{$\sigma$}	\\
		
		\hline
		
		H$\alpha$ & 6565 	& 	$0.898$ & $0.388$ & $0.842$ & $0.481$ & $0.803$	& $0.561$	& $-1.46$	& $0.961$	\\
		H$\beta$ & 4863 	& 	$0.882$ & $0.369$ & $0.840$ & $0.430$ & $0.795$	& $0.535$ 	& $-1.52$	& $0.854$ 	 \\
		\ion{S}{ii} & 6718 	& 	$0.839$ & $0.416$ & $0.773$ & $0.492$ & $0.798$	& $0.484$ 	& $-1.23$	& $0.832$	 \\
		\ion{S}{ii} & 6733 	&	$0.827$ & $0.433$ & $0.751$ & $0.516$ & $0.754$	& $0.527$ 	& $-1.36$	& $0.840$ 	 \\
		H$\gamma$ & 4342 	& 	$0.816$ & $0.418$ & $0.773$ & $0.465$ & $0.698$	& $0.772$ 	& $-2.81$	& $0.790$ 		\\
		\ion{O}{ii} & 3727 	& 	$0.749$ & $0.498$ & $0.700$ & $0.547$ & $0.556$	& $0.716$	& $-0.710$	& $0.817$ 		\\
		\ion{O}{iii} & 5008 & 	$0.743$ & $0.784$ & $0.673$ & $0.884$ & $0.673$	& $0.877$ 	& $-1.10$	& $1.268$ 		\\
		\ion{O}{iii} & 4960 &	$0.721$ & $0.773$ & $0.659$ & $0.858$ & $0.628$	& $0.890$ 	& $-1.35$	& $1.208$		\\
		\ion{N}{ii} & 6585 	& 	$0.680$ & $0.514$ & $0.677$ & $0.527$ & $0.411$	& $0.815$ 	& $-0.318$	& $0.757$		\\
		\ion{N}{ii} & 6550 	& 	$0.664$ & $0.570$ & $0.669$ & $0.579$ & $0.367$	& $0.880$	& $0.306$	& $0.820$ 		\\

		\hline
	\end{tabular}
	\caption{
Numerical properties of the various line reconstruction methods, for all 11 emission lines. The four methods are as follows: (1) from continuum principal components, fitting the 30 nearest neighbours, (2) from broad-band magnitudes, fitting the 30 nearest neighbours, (3) from continuum principal components, but instead of using the 30 nearest neighbours we used 30 random galaxies, and (4) from continuum principal components, but with a randomized sample (as a cross-test). For each reconstruction, we calculated the Pearson product-moment correlation coefficient $\rho$ and rms error $\sigma$. Emission lines are ordered by reconstructability using the first method.}
	\label{tab:ewfit}
\end{table*}

\subsection{Emission line reconstruction from broad-band magnitudes}
\label{sec:magfit}

To see if continuum PCA is any better than directly estimating emission lines from broad-band magnitudes, we performed the above analysis using the SDSS photometric magnitudes instead of the principal components. For this purpose, we used dereddened model magnitudes without any $K$-correction. The lack of $K$-correction is not supposed to significantly affect the procedure as the redshift distribution of the sampled galaxies is rather sharp.

While broad-band magnitudes are strongly correlated with continuum principal components, it is still interesting to see how lines are reconstructed from them. First of all, magnitudes are highly correlated with each other, whereas PCA eliminates covariance. Also, observed magnitudes are already `contaminated' with emission lines which might result in stronger correlations with EWs. Results of line reconstruction from magnitudes are plotted in Fig.~\ref{fig:magfit}. Compared with line reconstruction from PCA as plotted in Fig.~\ref{fig:ewfit}, no clear difference can be seen in terms of scatter, perhaps with the exception of more outliers being visible in the photometric case. Thus, log~EWs can be reconstructed from magnitudes almost as well as from the principal components. Quantitative results are listed in Col.~5-6 of Tab.~\ref{tab:ewfit}. We have to emphasize here that our sample contained strong emission line galaxies only, thus the strong correlation between magnitudes and log~EWs exists only for our sample and cannot be generalized to all galaxies.

\subsection{Non-local line reconstruction from the continuum}

To test whether a single global linear model is sufficient to reproduce the lines, we repeated the procedure \revision{of} estimating log~EWs from the continuum principal components as described in Sec.~\ref{sec:pcafit} but instead of using the 30 nearest neighbour galaxies, we randomly selected 30 galaxies from the entire sample. Another difference was that the $\chi^2$ of the fit was not weighted by the inverse-square of the distance from the query point to relax the effect of locality. By looking at Col.~7-8 of Tab.~\ref{tab:ewfit}, it is somewhat surprising that correlation coefficients and rms errors of the individual lines did not get much worse. By looking at panel (c) of Fig.~\ref{fig:bpt_random}, one can clearly see, however, that that the star-forming branch of the BPT diagram cannot be reconstructed this way, and the AGN sequence is also greatly distorted. The conclusion is that emission line log~EWs cannot be explained by a simple, global linear relationship with continuum principal components. Thus, local fitting from nearest neighbours is necessary to reconstruct the BPT from either continuum principal components or broad-band magnitudes.

\subsection{Cross-tests with randomized data}
\label{sec:randomfit}

As another test, we shuffled the sample and randomly paired continuum principal components with emission line vectors of other galaxies to break the locality in the PCA space. Results are listed in Col.~9-10 of Tab.~\ref{tab:ewfit}. It is not surprising that correlations almost entirely disappear in the shuffled case (note that $\rho$ values are multiplied by $10^{2}$). This supports that information about the emission lines is indeed encoded in the continuum of a galaxy spectrum. 

\subsection{Reconstructing the BPT}

\begin{figure}
	\includegraphics{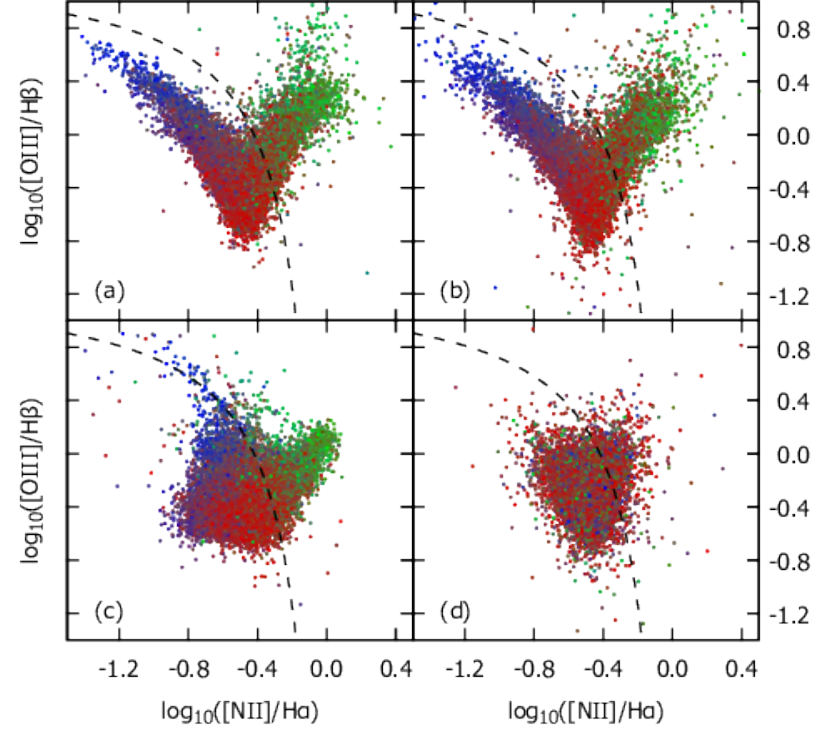}
	\caption{BPT diagrams with log~EWs reconstructed from the stellar continuum using different methods and cross-tests. Panel (a) shows reconstructed log~EWs from continuum principal components using local linear regression from the 30 nearest neighbours. Panel (b) is the reconstruction of lines from broad-band magnitudes by local linear regression from the 30 nearest neighbours. Panel (c) was drawn from lines estimated using the global linear regression technique from 30 randomly selected galaxies. Panel (d) is the cross-test using local linear regression but with shuffled continuum principal components. The colour coding of the data points is based on the original BPT as in panel (a) of Fig.~\ref{fig:bpt_classes}. See the text for discussion.}
	\label{fig:bpt_random}
\end{figure}

As the BPT diagram is generally used to classify emission line galaxies into star-forming and AGN, it is very informative to see how well the various methods can recover it solely from the continuum. The difference among the line estimation methods introduced above is obvious once the BPT diagram is plotted from the reconstructed log~EWs, as it was done in Fig.~\ref{fig:bpt_random}. Local linear fitting of EWs using the nearest neighbours and reconstructing lines from broad-band magnitudes give similarly fair, qualitatively correct BPT diagrams while the randomization of the sample disrupts the diagram entirely.

To further analyse the properties of a reconstructed BPT diagram, we will stick to local linear regression based on the continuum principal components. In Fig.~\ref{fig:bpt_classes}, we plot the original BPT for reference, the reconstructed diagram for all galaxies, and two diagrams showing AGNs and star-forming galaxies only, as classified by \citet{Kauffmann2003a}.

The first thing to see in panels~(c)~and~(d) of Fig.~\ref{fig:bpt_classes} is the mixing of weak star-forming galaxies with weak AGNs in the bottom corner of the reconstructed BPT diagram. The mixing is caused by the bad reconstructability of the [\ion{O}{iii}] line which is most likely due to lack of a strong correlation between AGN activity level and the stellar continuum.

\section{Revisiting star-forming/AGN separation}
\label{sec:svm}

The mixed, low activity -- low star formation rate region is located at the bottom corner of the [\ion{N}{ii}]/H$\alpha$--[\ion{O}{iii}]/H$\beta$ BPT diagram. Empirically drawn BPT diagrams are noisy enough to smear pure star-forming galaxies and mixed star-forming/AGNs together in this part of the BPT so it is an interesting question whether it is possible to segregate galaxies into two distinct classes or not by incorporating information on the stellar continuum into classification model. By visually inspecting the projections of the 5D continuum PCA space, one can see that while AGNs and star-forming galaxies occupy different loci, they cannot be clearly separated into two disjoint sets by cuts in any principal component dimensions, nor the distribution of galaxies is bimodal. We turned to SVM, a machine learning algorithm, to determine an empirical segregation plane between the two classes in the continuum-PCA--line-PCA space.

\subsection{Support vector machines}

SVM are supervised learning algorithms which can be trained to automatically classify multidimensional data vectors into two disjoint sets \citep{svm, SVM_R}. The training phase starts with compiling a \textit{training set} of data vectors that are tagged as either belonging to class $A$ or class $B$. During learning, the model will find a hyperplane in the space of data vectors which separates the elements of $A$ and $B$ with the largest possible margin\footnote{Since this is often not possible in the original space of data vectors because the distributions of the two halves of the training set are non-convex, kernel functions are used to map training set vectors into a higher dimensional space where linear segregation is possible \citep{SVM_kernels}. Another option to handle non-convex situations is to find a \textit{best possible} segregation plane which minimizes the overlap.}. Once the model is trained, it can be used to classify any query point into one of the two classes.

\subsection{Automatic star-forming/AGN classification}

\begin{figure}
	\includegraphics{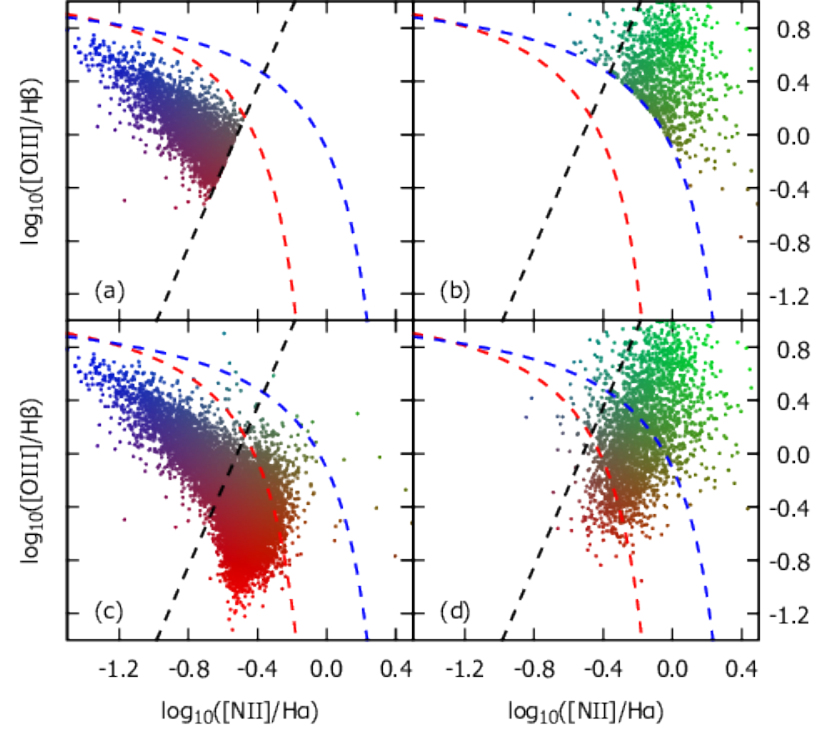}
	\caption{BPT diagrams resulting from the SVM-based star-forming/AGN separation. Panels (a) and (b) show star-forming and AGN galaxies used to train the algorithm. Panels (c) and (d) display the outcome of the automatic classification. The theoretical segregation line of \citet{Kewley2001} is drawn in blue and the empirical one of \citet{Kauffmann2003a} in red. Our star-forming/low-activity line is in black. Colour coding of the data points is based on the original BPT as in panel (a) of Fig.~\ref{fig:bpt_classes}. See the text for discussion.}
	\label{fig:bpt_svm}		
\end{figure}

We compiled the training set from our emission line galaxy sample by selecting galaxies on the BPT that could be classified with high confidence either as pure star-forming or AGN. To select high-confidence AGNs only, we picked galaxies above the theoretical maximum starburst line of \citet{Kewley2001}:
\begin{equation*}
	\log_{10}\left(\frac{[\ion{O}{iii}]}{H\beta}\right) > 0.61 \left[ \log_{10}\left(\frac{[\ion{N}{ii}]}{H\alpha}\right)-0.47 \right]^{-1} + 1.19.
\end{equation*}
Star-forming galaxies were selected to fall below the empirical starburst line of \citet{Kauffmann2003a}:
\begin{equation*}
	\log_{10}\left(\frac{[\ion{O}{iii}]}{H\beta}\right) < 0.61 \left[ \log_{10}\left(\frac{[\ion{N}{ii}]}{H\alpha}\right)-0.05 \right]^{-1} + 1.3,
\end{equation*}
and at the same time be above the \revision{following line, defined by us:}
\begin{equation*}
	\log_{10}\left(\frac{[\ion{O}{iii}]}{H\beta}\right) > 3 \log_{10}\left(\frac{[\ion{N}{ii}]}{H\alpha}\right) + 1.55.
\end{equation*}
\revision{The line was drawn empirically to cut out the most reliably identifiable part of the star-forming population.} Curves on Fig.~\ref{fig:bpt_svm} illustrate these cuts.

We used the first five continuum principal components and the first four log~EW principal components of the training set galaxies as input data vectors to SVM. By combining information from the continuum into the training, we might hope a better separation of the two galaxy types in the mixed lower corner of the BPT than simply from the emission lines. As SVM is a strictly empirical model, we shall not, however, draw far-reaching theoretical conclusions from its outcome. Since our training set was not containing the mixed region, it was directly separable into two disjoint classes by a linear cut. Consequently, data points did not need to be projected into any higher dimensional space by a kernel function, like in most applications of SVM, we simply ran it on the $5+4$ dimensional vectors of the continuum + line PCA space.

We plot the results of the SVM-based classification in Fig.~\ref{fig:bpt_svm}. Panels (a) and (b) show the two training set classes (star-forming and AGN, respectively) with log~EWs of the originally measured emission lines. Panels (c) and (d) show the outcome of the SVM classification. Even though the mixed region was not included in the training set at all, SVM reproduced the empirical segregation line of \citet{Kauffmann2003a} surprisingly well, with only about $6$ per cent of the sample scattered into the opposite region.

\section{Generating realistic random emission lines}
\label{sec:recipe}

\subsection{\revision{The stochastic recipe}}
Based on our findings, we propose a simple stochastic recipe to generate \revision{a realistic distribution of} emission lines for stellar population synthesis models that provide the continuum only. The algorithm works by expressing the model continuum as a linear combination of the basis vectors derived from PCA of the continua of SDSS galaxies. According to these principal components, the model spectrum is classified into one of the $60$ continuum classes. We used $k$-means clustering to define the continuum classes, as described in Sec.~\ref{sec:kmeans}.

Let us denote the average continuum vector with $e_{0,\lambda}$ and the PCA basis vectors with $e_{i,\lambda}$, where $i$ indexes the five dimensions of the PCA space and $\lambda$ goes over the wavelength bins. Continuum classes are given by the centre of mass vectors $c_{n, i}$ where $n$ indexes the $60$ classes. Within each class, model lines are randomly generated from a multivariate Gaussian distribution. The mean line log~EWs $\vecvar{m}_n$ and the covariance matrices $\vecvar{C}_n$ of the distributions are pre-calculated from the real galaxy sample and provided for each of the $60$ continuum classes.

The detailed recipe for generating realistic emission lines given a stellar continuum model spectrum is the following.
\begin{enumerate}
	\item Rebin the rest-frame model spectrum $s_\lambda$ to the grid of the basis vectors and normalize it as described in Sec.~\ref{sec:contpca} to get $\tilde{s}_\lambda$.
	\item Subtract the average continuum $e_{0,\lambda}$ from the normalized spectrum.
	\item Express the continuum as a linear combination of the provided basis by calculating the dot products $a_i = \sum_\lambda{\left[ e_{i, \lambda} \cdot (\tilde{s}_\lambda - e_{0,\lambda}) \right] }$
	\item Find the class centre $c_{n, i}$ in the continuum PCA space that is the closest (in Euclidean distance) to the vector $a_i$ of the linear coefficients.
	\item Take the covariance matrix $\vecvar{C}_n$ and mean line log~EW vector $\vecvar{m}_n$ of the line distribution within the closest class and generate a random vector of line log~EWs from the corresponding multivariate Gaussian distribution.
\end{enumerate}

Data necessary to generate random lines are published on the paper's web site.

\subsection{$k$-means clustering}

$k$-means clustering is a machine learning algorithm that classifies data points based on their distances from cluster centres: points belonging to a cluster must be closer to the centre of mass of that particular cluster than any other clusters'\footnote{The $k$-means algorithm basically constructs a Voronoi tessellation from the data vectors, with seeds being the centres of mass of the clusters.}. This implicit definition of a cluster makes finding the best exact solution a hard problem, but heuristic, randomized algorithms exist that can find a reasonable clustering relatively fast \citep{Forgy1965, MacQueen1967, Hartigan1979, Lloyd1982}. The only inputs of $k$-means clustering are the data vectors and $k$, the number of clusters wanted. The output is the centres of mass of the $k$ clusters. Once the latter are known, new points can be classified simply by measuring their distances from the cluster centres and putting them into the one with the closest centre.

\subsection{Automatic classification of emission line galaxies}
\label{sec:kmeans}

To construct our stochastic model of emission lines, we started from the $5 + 4$-dimensional vector space of continuum and log~EW principal components of our high signal-to-noise ratio SDSS galaxy sample. First, we classified galaxies into continuum--log~EW classes using the $k$-means clustering function of R and the algorithm of \citet{MacQueen1967}.

\begin{figure}
	\includegraphics{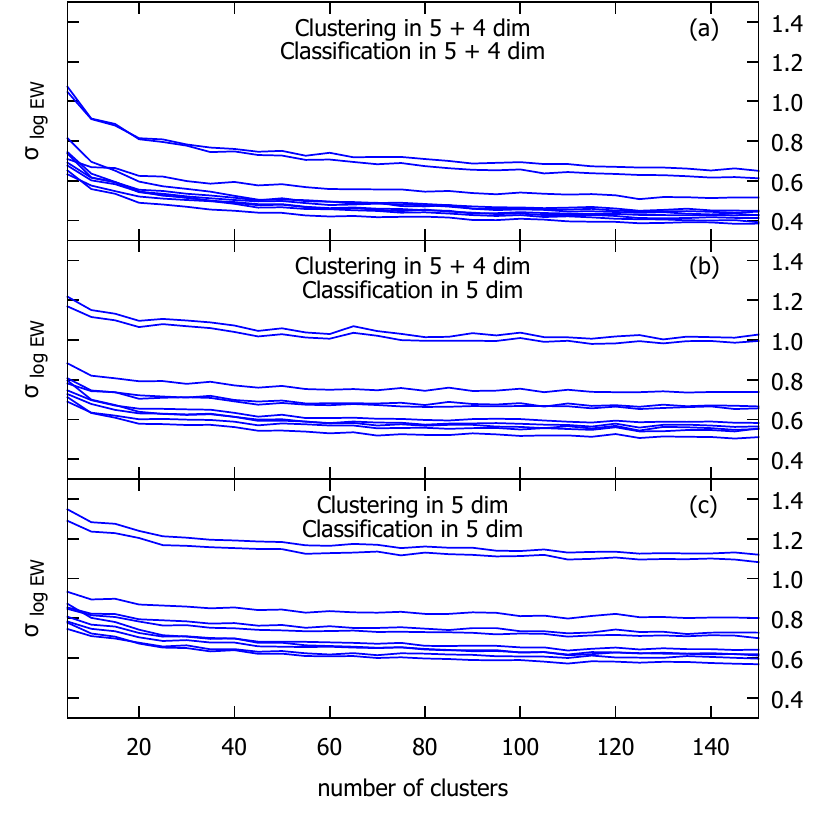}
	\caption{Variance of log~EW of the 11 strong emission lines, averaged over all clusters, as a function of the number of clusters. Panel (a) refers to the case of considering both the continuum and line principal components for clustering and classification. Variance results from the sum of line measurement errors and uncertainty due to the finite size of the clusters. Panel (b) shows the effect of misclassification when only the continuum principal components are used to classify galaxies, with the clustering done in both the continuum and line log~EW PCA space. Misclassification will add extra scatter to the randomly generated log~EWs, cf. Sec.~\ref{sec:recipeproblems} and Fig.~\ref{fig:bpt_recipe}. Panel (c) illustrates the case of performing both the clustering and classification in the continuum principal component space only. Even with the additional variance due to misclassification, using both the continuum and line log~EW principal components for the clustering is favourable.}
	\label{fig:variance}
\end{figure}

To choose the right number of clusters, one has to consider the variance of emission line log~EWs as functions of the number of the clusters. The variance in each cluster is supposed to be decreasing as the number of clusters is growing since clusters are becoming smaller. The minimum variance is limited by the noise in the data. The $\sigma(k)$ curves for all emission lines are plotted in panel (a) of Fig.~\ref{fig:variance}. Hence, to minimize the variance of line strengths within each class, we chose $k = 60$ as all curves get essentially flat above this value. \revision{For training sets of different sizes and characteristics, a similar analysis of the variance is advisable to determine the input parameter of $k$-means clustering.}

\subsection{Modelling the emission line distributions}
\label{sec:ewdist}
Once $k$-means clusters in the continuum-PCA--line-PCA space are determined in the way described in Sec.~\ref{sec:kmeans}, we have to model the distribution of emission line log~EWs within each cluster. If the number of clusters is sufficiently high, clusters will become small enough that the distribution of emission lines within them can be well modelled by a multivariate Gaussian distribution parametrized with $\vecvar{m}_n$ and $\vecvar{C}_n$. \revision{We note that a multivariate Gaussian distribution, when its entire covariance matrix is known, does not only account for individual line strengths but also for line ratios, including ratios from the same line series.} It is also important to mention that, while we did the $k$-means classification of galaxies in the $5 + 4$-dimensional continuum + log~EW space, stellar population synthesis models yield the continuum coefficients only. As a result, when classifying model continua, we measure distances from cluster centres in the 5D continuum-PCA subspace only. This will introduce some mixing among clusters as determined by $k$-means and cause somewhat larger scatter in the randomly generated log~EWs than what it would be based solely on the $\vecvar{C}_n$ covariance matrices. This effect is shown in panel (b) of Fig.~\ref{fig:variance}. It is still worth using the entire $5 + 4$-dimensional space to run the $k$-mean classification because the resulting variances are still lower than using the continuum principal components only, cf. panel (c) of Fig.~\ref{fig:variance}. Also, because the covariance of the lines is treated stochastically, there will be random scatter in line ratios as well.

\begin{figure}
	\includegraphics{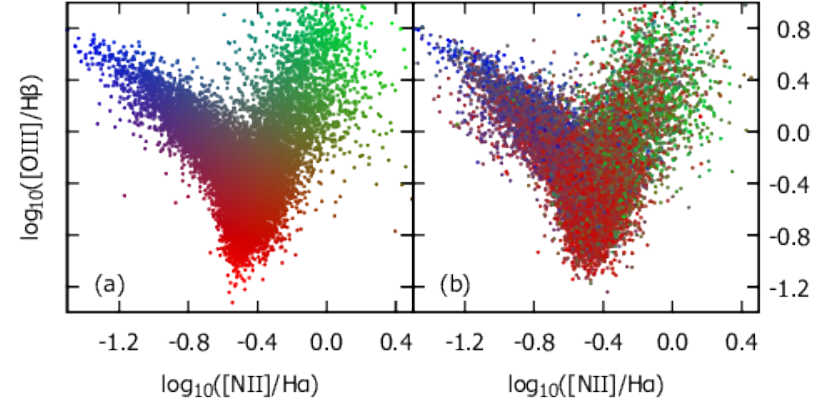}
	\caption{Panel (a) shows the original diagram plotted from the directly measured emission line for reference. Panel (b) is the BPT diagram plotted from stochastically generated emission lines based on the recipe described in Sec.~\ref{sec:recipe}. While lines were generated from a multivariate distribution randomly, based on the location of the continuum in the PCA space, the resulting BPT diagram resembles the original one remarkably well, although more scatter and significant mixing of galaxy types is visible. Colour coding of the data points is based on the original BPT as in panel (a) of Fig.~\ref{fig:bpt_classes}.}
	\label{fig:bpt_recipe}
\end{figure}

A direct test of the algorithm is to take the galaxies of our SDSS sample,  generate emission lines based on their fitted continua and see how well the BPT diagram can be reconstructed. Results of this procedure are plotted in Fig.~\ref{fig:bpt_recipe} where we also show the original BPT for reference in panel~(a) next to the stochastically generated BPT in panel~(b). While the curve of star-forming galaxies and the AGN mixing sequence is reproduced reasonably well, there \revision{are} also a large number of red data points corresponding to the bottom corner of the original BPT visible in all regions of the plot.

\subsection{Shortcomings of the method}
\label{sec:recipeproblems}

The recipe outlined above yields line EWs only, and is sufficient to reproduce the flux excess caused by emission lines but not line widths. In general, line widths should be taken to be equal to the velocity dispersion, at least in the case of star-forming galaxies. Width distributions of broad lines would need to be investigated to generate AGN lines with realistic breadth distribution.

\revision{As we pointed out in Sec.~\ref{sec:ewdist}, using only the continuum principal components to generate log~EWs introduces additional variance due to the mixing of the classes as defined in the continuum + line space. Additionally, as lines are randomly generated based on a multivariate Gaussian distribution, EWs and line ratios are not guaranteed to be correct for individual galaxies, but will be for the entire ensemble of mock galaxies.}
 
\revision{If the goal is to generate realistic emission lines for individual model continua, we suggest using the local linear regression method as described in Sec.~\ref{sec:pcafit}. While that technique yields more accurate emission line estimates, it also requires a much larger input data set and more heavyweight algorithms.}

\section{Summary and future work} 
\label{sec:summary}

We have measured the emission lines of galaxies from SDSS~DR7 to analyse the correlation between the emission lines and the stellar continua in the optical wavelength range. We have developed an algorithm, noise limited fitting, to accurately measure the parameters of broad and asymmetric emission lines, yet avoid overfitting of narrow, symmetric lines. We have also demonstrated how to correct for discrepancies between theoretical stellar continuum modelling and real measurements by low-pass filtering the residual before emission line fitting.

In Sec.~\ref{sec:emlines}, we have shown that optical emission line log~EWs can be reasonably well reconstructed from both the optical stellar continuum and broad-band magnitudes of galaxies. 

\revision{The main practical use case of our method is to generate emission lines for stellar continua from stellar population synthesis models, provided that the models fall into the wavelength and physical parameter coverage of our training set. Since our sample contained strong emission line galaxies only (with all $11$ prominent lines measured), the results cannot be generalized to any type of galaxy without extending the training set, but the algorithm still applies. Also, further research is necessary to use our line reconstruction method for galaxies with fewer and weaker lines: correlations between the stellar continuum and the probability of the very presence of weak emission lines need to be taken into account.}

\revision{Another application of our method is to estimate emission lines of photometric galaxies. The technique readily works for the SDSS \textit{ugriz} filter set, but the existing training set can be adapted to other filter systems as well. While one simple way to do this is to compute synthetic photometry from the spectra, building a new training set by cross-matching the photometric measurements made with the other filter set to our spectroscopic sample is a better option (provided that the survey overlaps with the SDSS), since it would automatically account for the unknown systematics in spectrophotometry. With the outlined modifications, our technique will be of great value for analysing data from large photometric surveys like PanSTARRS and the LSST.} 

\revision{Additionally, by correcting for the contributions of strong emission lines to broad-band magnitudes, our method can be useful in improving template-based photometric redshift estimation algorithms to narrow the performance gap between the theoretical and the empirical approach.}

In Sec.~\ref{sec:svm}, we used a supervised machine learning algorithm, SVM, to verify the empirical demarcation line between star-forming galaxies and AGNs defined by \citet{Kauffmann2003a}. Even though we used only extreme starburst galaxies and strong AGNs to train the algorithm, SVM yielded a result very similar to the analytical segregation curve, only about 6~per~cent of galaxies in the bottom corner of the BPT diagram got classified differently. A future application to SVM would be to revisit the Seyfert/LINER separation as it was done in \citet{Kewley2006}.

Finally, in Sec.~\ref{sec:recipe}, we gave a very simple recipe to generate random emission lines with realistic EWs on top of stellar continua generated by stellar population synthesis modes. We have demonstrated that, despite its simplicity, the method can qualitatively reconstruct the BPT. Our model has its application when the objective is not the accurate modelling of \revision{the emission lines of individual galaxies} but rather generating stochastic mock catalogues with more realistic broad-band magnitudes.

\section*{Acknowledgements}

The realization of this work was supported by the Hungarian OTKA NN grants 103244 and 114560.

\bibliographystyle{mn2e}

\bsp

\label{lastpage}

\end{document}